\documentclass[11pt,twoside]{article}
\usepackage{apn3conf}

\begin{document}

\contact{Rachel Deacon}
\email{rdeacon@physics.usyd.edu.au}

\title{Radio Observations of Masers in Post-AGB Stars}

\author{Rachel M. Deacon \altaffilmark{1} \& Anne J. Green} 
\affil{School of Physics A29, University of Sydney, NSW Australia 2006}

\author{Jessica M. Chapman} 
\affil{CSIRO Australia Telescope National Facility, PO Box 76, Epping NSW Australia 1710}

\altaffiltext{1}{Affiliated with the Australia Telescope National Facility, CSIRO}

\begin{abstract}
We present observations of 86 post-Asymptotic Giant Branch (post-AGB) stars of OH maser transitions, taken with the Parkes Telescope between September 2002 and August 2003.  Post-AGB stars are the precursors of planetary nebulae, which have a wide range of morphologies that are not well explained.  By studying the circumstellar envelopes of post-AGB stars through the masers produced in them, we hope to shed light on the origin of planetary nebula morphologies.
\end{abstract}

\paindex{Deacon, R. M.}
\aindex{Green, A. J.}
\aindex{Chapman, J. M.}

\titlemark{Radio Observations of Masers in Post-AGB Stars}
\authormark{Deacon, Green \& Chapman}


\section{Introduction}

\addtocounter{footnote}{1}

Planetary nebulae (PN) are spectacularly beautiful and diverse in form. While some appear circular, others have the shapes of butterfly wings and show elliptical or bipolar shapes, in some cases with complex, filamentary structures \htmladdnormallinkfoot{hubble}{http://hubblesite.org/newscenter/archive/category/nebula/planetary/}. One possible cause of the variations is that magnetic fields from the stars constrain the flow of stellar winds in different ways. Alternatively, companion stars or planets may produce gravitational effects and rotation of the central stars may be an important factor. 

The different geometries seen in PN are also evident in post-AGB stars and it seems likely that the shaping of non-spherical PN winds begins early in the post-AGB phase. By studying the post-AGB stars we may be able to determine what causes the complicated structures seen in some PN. 

\section{Selection of Source Sample}

A sample of 86 post-AGB stars has been selected from a previous study of OH 1612 MHz maser sources in the Galactic plane (Sevenster et al. 2001). Sevenster (2002a,b) has compared the far-infrared IRAS (InfraRed Astronomical Satellite) and MSX (Mid-course Space Experiment) properties of these sources and found that the far-infrared colours provide a powerful tool for distinguishing between early and late, and low and high  mass, post-AGB objects.  

The present project is to observe maser emission from hydroxyl (OH), water (H$_2$O) and silicon monoxide (SiO) molecules that are located in the outflowing stellar winds. Each molecule exists in different physical conditions and the maser emission from these molecules is produced at different locations within the circumstellar winds, providing information on the wind velocities, and on whether the winds are likely to be spherically symmetrical or distorted. 

Sevenster has argued that the more massive sources in the sample, designated 'LI' (Sevenster 2002b) are likely precursors of bipolar PN while the less massive sources ('RI') evolve into elliptical or spherical PN.  Analysis of the maser properties of LI and RI objects, together with a study of the early and late post-AGB objects, will give insights into the origins of the different morphologies in PN.  

\section{Observations}

To date we have made observations in the four ground-state OH maser transitions at 1612, 1665, 1667 and 1720 MHz for all the sources in the sample, using the Parkes radio telescope (R. M. Deacon, J. M. Chapman, \& A. J. Green, in preparation). After Hanning smoothing, the final velocity resolution of the spectra was 0.18 km s$^{-1}$ per channel. The average rms noise on a single channel was 70 mJy beam$^{-1}$.  The angular resolution achieved was $12.6'$.  Typically, each source was observed for 10 mins.

Higher-resolution observations of sources with 1720 MHz detections were carried out in August 2003 with the Australia Telescope Compact Array (ATCA) particularly to resolve confused sources.  

\section{Results}

Figure 1 shows examples of maser spectra for sources in our sample. We have classified them in six different categories, as follows. Many exhibit a double-peaked spectrum ({\bf D}) with steep outer edges and sloping inner edges between the two peaks. This spectral profile is characteristic of an expanding spherical shell, with the strongest emission coming from the region of the shell closest and furthest from the observer along the line of sight (Reid et al. 1977). 

We see several variations on the classic double-peaked profiles. The {\bf De} spectra are very asymmetric showing that the maser emission is much stronger on one side of the star than on the other. The {\bf Dw} stars have spectra with sloping outer edges as well as sloping inner edges. These are expected to be stars with bipolar shells. More unusual is the {\bf DD} source with four emission peaks. At present there are only two sources known with profiles like this, and one of them is in our sample.  We also find a small number of {\bf S} sources. These show only a single peak of maser emission, but otherwise have characteristics in common with AGB and post-AGB stars. Finally the {\bf I} or irregular spectra have multiple emission peaks and an unusually large velocity dispersion. Post-AGB stars with irregular spectra have previously been associated with exotic envelope geometries (Zijlstra et al. 2001)

\begin{figure}[ht!]
\epsscale{0.84}
\plotone{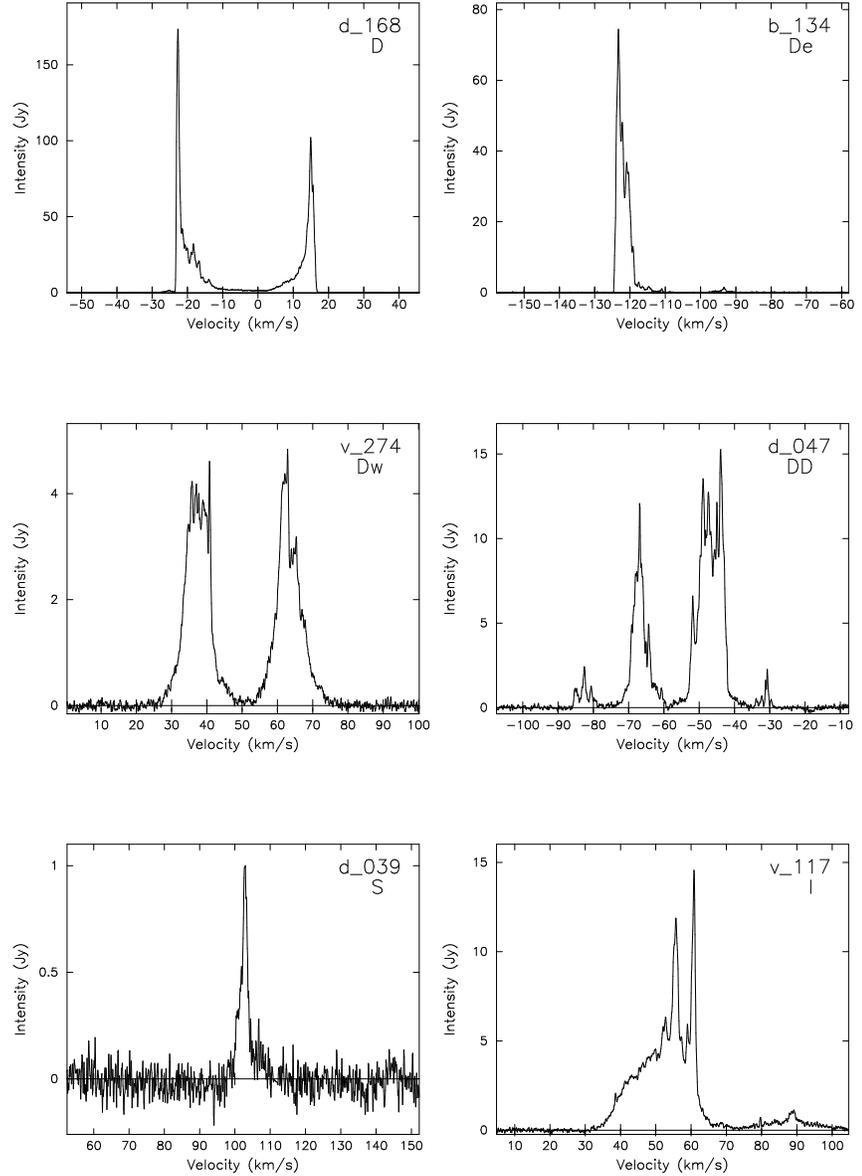}
\caption{Examples of spectra from each classification category.  The catalogue name (Sevenster 2001) of each star is given in the top right-hand corner with the category type given below the name. \label{Fi:examples}}
\end{figure}

OH 1720 MHz maser emission was detected for one source in the sample - b292 - confirming a previous measurement (Sevenster \& Chapman 2001). 

Table \ref{Ta:dets_by_profile} lists the number of sources detected in the OH 1612, 1665 and 1667 MHz lines for each profile type. The fraction of sources detected at 1665 and 1667 MHz are approximately 30\% and 50\% respectively. One significant result is that 63\% of the 1665 MHz detections are classified as {\bf I}, in contrast to 1612 (13\% {\bf I}) and 1667 MHz  (16\% {\bf I}). 

\begin{deluxetable}{c c c c}
\tablewidth{0pt}
\tablecaption{Number of detections by frequency and spectral profile. \label{Ta:dets_by_profile}}
\tablehead{
\colhead{Profile class} & \colhead{1612 MHz} & \colhead{1665 MHz} & \colhead{1667 MHz}}
\startdata
{\bf D} & 60 & 5 & 25 \\
{\bf DD} & 1 & 0 & 0 \\
{\bf De} & 4 & 0 & 2 \\
{\bf Dw} & 6 & 1 & 1 \\
{\bf I} & 11 & 17 & 7 \\
{\bf S} & 4 & 4 & 9 \\[1pt]
\tableline \\[-9pt]
Total Sample & {\bf 86} & {\bf 27} & {\bf 44} \\[-2pt]
\enddata
\end{deluxetable}
\section{Conclusions}

The aim of this project is to determine how the maser characteristics correlate with trends in the infrared colours of the sample.  Preliminary results show a lower number of LI objects detected at 1667 and (particularly) 1665 MHz, compared with the other sub-groups in the sample.  This shows a possible correlation between mainline emission and stellar mass in post-AGB stars, with mainline masers being preferentially disrupted by the bipolar outflows and/or magnetic fields in higher-mass PN precursors.  

Ongoing observations include the H$_2$O and SiO masers in these stars, polarisation observations to study magnetic fields, and high-resolution imaging of the more unusual sources. 

\acknowledgments

The Parkes Telescope and ATCA are funded by the Commonwealth of Australia for operation as National Facilities managed by CSIRO. Rachel Deacon acknowledges the receipt of an Australian Postgraduate Award.

\end{document}